\pdfoutput=1

\let\singlecol\undefined 

\ifdefined\singlecol
  \documentclass[useAMS,usenatbib,referee]{aa}
\else
  \documentclass[useAMS,usenatbib]{aa}
\fi

\usepackage[pdftex]{}
\usepackage{times}
\usepackage{wasysym}
\usepackage{url}
\usepackage{lineno}
\usepackage{amssymb}
\usepackage{subfigure}
\usepackage{amsmath}
\usepackage{siunitx}
\usepackage{xcolor}

\let\boldcorrections\undefined

\ifdefined\boldcorrections
  
\else
  
\fi

\usepackage{hyperref}
\hypersetup{
    colorlinks=true,
    linkcolor=magenta,
    citecolor=blue,
    filecolor=magenta,      
    urlcolor=cyan,
}

\newcommand{\etacar}{$\eta$~Car}
\newcommand{\etacare}{$\eta$~Carinae~}

\newcommand{\UNITS}[1]{\,\mathrm{#1}}

\newcommand{\grad}{$^{\circ}$}

\newcommand{\fermi}{\emph{Fermi}-LAT}

\newcommand{\hess}{H.E.S.S.}

\def\apjl{ApJ}%
\def\apjs{ApJS}%
\def\ao{Appl.~Opt.}%
\def\aap{A\&A}%
\addto\extrasenglish{%

}

\DeclareSIUnit\solarmass{\ensuremath{M_\odot}}
\DeclareSIUnit\solarradius{\ensuremath{R_\odot}}
\DeclareSIUnit\year{yr}
\DeclareSIUnit\solarluminosity{\ensuremath{L_\odot}}
\DeclareSIUnit\gauss{G}
\DeclareSIUnit\erg{erg}
\DeclareSIUnit\pc{pc}
\def\@fnsymbol#1{\ensuremath{\ifcase#1\or *\or \dagger\or \ddagger\or
   \mathsection\or \mathparagraph\or \|\or **\or \dagger\dagger
   \or \ddagger\ddagger \else\@ctrerr\fi}}

\defcitealias{White2020}{W20}
\newcommand{\RW}{\citetalias{White2020}}
\defcitealias{Ohm2015}{O15}
\newcommand{\SO}{\citetalias{Ohm2015}}

\begin{document} 

\title{Probing cosmic ray escape from \(\eta\) Carinae}
\author{S.~Steinmassl \inst{1}  \and
        M.~Breuhaus \inst{1} \and
        R.~White \inst{1}  \and
        B. Reville \inst{1}
        \and J.A.~Hinton \inst{1} 
        }

\institute{Max-Planck-Institut f\"ur Kernphysik, Postfach 103980, D 69029 Heidelberg, Germany}

\date{Accepted 2023-09-06. Received 2023-xx-xx; in original form 2023-xx-xx}
  
\abstract
{
The binary stellar system $\eta$ Carinae is one of very few established astrophysical hadron accelerators. It seems likely that at least some fraction of the accelerated particles escape from the system. Copious target material for hadronic interactions and associated $\gamma$-ray emission exists on a wide range of spatial scales outside the binary system. This material creates a unique opportunity to trace the propagation of particles into the interstellar medium.
Here we analyse $\gamma$-ray data from \fermi\ of \etacare and surrounding molecular clouds and investigate the many different scales on which escaping particles may interact and produce $\gamma$-rays. 
We find that interactions of escaping cosmic rays from \etacare in the wind region and the Homunculus Nebula could produce a significant contribution to the $\gamma$-ray emission associated with the system.
 Furthermore, we detect excess emission from the surrounding molecular clouds.
The derived radial cosmic-ray excess profile is consistent with a steady injection of cosmic rays by a central source. However, this would require a higher flux of escaping cosmic rays from  $\eta$ Carinae than provided by our model. Therefore it is likely that additional cosmic ray sources contribute to the hadronic $\gamma$-ray emission from the clouds.
}


\maketitle

\section{Introduction}

Colliding wind binaries (CWBs) are a well-established class of Galactic particle accelerator, with non-thermal emission seen from a number of systems~\citep{1989Natur.340..449M, 1992ApJ...393..329C, 1999ApJ...515..762C,2000MNRAS.319.1005D, Marcote21,Palacio23}. In $\gamma$-rays, only two systems have been detected so far: $\eta$ Carinae \cite[e.g.][]{Fermi:1yr} and $\gamma^2$ Velorum \citep{Marti-Devesa_et_al_2020_gamma_velorum}, located $(2.35 \pm 0.05)$~kpc \citep{Smith_2006} and $(0.34 \pm 0.01)$~kpc \citep{North2007} from Earth, respectively. $\eta$ Carinae (henceforth \etacar) is the more luminous of these two systems and has therefore been studied in greater detail. 
Over its 15-year lifetime, the Fermi Large Area Telescope (LAT) has probed three periastron passages of this highly eccentric \cite[$\epsilon \approx 0.9$,][]{Mehner_et_al2015} CWB with an orbital period of $\sim$5.5~yr \citep[e.g.][]{Damineli2008, Teodoro2016}. The high eccentricity and relatively short orbital period, combined with its dense matter and photon fields make \etacare a unique source in which to study high-energy astrophysical processes in a time-dependent environment. 

The Astro-Rivelatore Gamma a Immagini Leggero (AGILE) satellite was used to detect \etacar\ for the first time in $\gamma$-rays \citep{Tavani2009}. The detection was soon confirmed by observations with the \fermi\ satellite \citep{Fermi:1yr, Fermi:2yr}. Since then, several works have analysed the $\gamma$-ray emission measured by \fermi\
\citep{EtaCar:Fermi10, EtaCar:Farnier11, Reitberger2012, Reitberger2015, Balbo2017, White2020, MartiDevesa21, 2022ApJ...933..204A}. Two
components are apparent in the emitted spectrum above and below \SI{\sim 10}{\giga\electronvolt} \citep[see e.g.][]{EtaCar:Fermi10, EtaCar:Farnier11}. During the periastron phase when the stars are closest to each other, the emission increases. The peak emission at \si{\giga\electronvolt} energies around periastron increases by \SI{\sim 40}{\percent} relative to the baseline flux \citep{MartiDevesa21}. Emission up to 400~GeV has now been established using the High Energy Stereoscopic System (\hess)~\citep{HESS_2020_EtaCar}.

The $\gamma$-ray emission from \etacar\ is generally believed to be emitted by cosmic rays (CRs) accelerated by the shocks that enclose the wind collision region (WCR), where the supersonic winds from the two stars collide to form a pair of standing shocks, separated by a contact discontinuity.
The shocks are approximately located at the surface (cap) where the ram pressures from both winds are equal. At the shocks, electrons and protons can be accelerated to relativistic energies and emit $\gamma$-rays via synchrotron and inverse Compton processes or via proton-proton collisions with subsequent $\pi^0$-decay, respectively. For the latter process a dense target material is needed to produce a detectable gamma-ray signal.
\citet{Bednarek11} were the first to point out that the two shocks in the system should have distinct properties due to the different stellar mass loss rates and terminal wind velocities. These properties result in two different accelerated particle populations. These ideas were developed further by \citet{Ohm2015} and \citet{White2020} (henceforth \SO\ and \RW\ respectively) where models were constructed to account for the variable $\gamma$-ray emission observed by \fermi\ and the non-thermal X-ray emission detected with the NuSTAR satellite \citep{Hamaguchi2018}.

In \RW, a non-negligible amount of the accelerated protons escape the WCR. Such particles would naturally interact with the surrounding material further out. 

The complex environment around \etacar\ provides many potential interaction regions. 
Though the inner structures, including the binary system and its immediate vicinity, can not be resolved by \fermi\ or \hess, 
the nearby molecular clouds in the Carina Nebula Complex (CNC) and nearby Gum\,31   
are resolvable. 
Significant emission associated with these clouds, above the expectations for the `sea' of diffuse CRs, was reported in \RW. Here we explore this emission in more detail. 
 
 Recently, \cite{Ge_et_al_2022_clouds_around_EtaCar} published an analysis of the excess $\gamma$-ray emission around \etacar\ considering two Gaussian regions, for which significant emission, likely of hadronic origin was found. They conclude that the emission could be connected to young massive stellar clusters in the region, such as Trumpler 14 and 16. Nevertheless, the authors were not able to rule out, that \etacar\ or yet unknown CR sources are the acceleration sites of the particles producing the $\gamma$-ray emission.  

In this article, we follow up on these ideas and perform a detailed analysis of the emission from \etacar\ and the surrounding excess emission considering not only two but four different regions. These regions correspond to specific nearby molecular cloud structures connected to the Carina Nebula-Gum 31 complex. Additionally, we analyse possible $\gamma$-ray emission from escaping CRs interacting with material on small and large scales around \etacar.

While this work focuses primarily on escaping particles accelerated in the WCR, it is also possible that particles have been accelerated in the blast wave from the great eruption of 1843, when the system released more than \SI{10}{\solarmass} of material \citep{Smith_2006} with high velocity
\citep{Ohm_et_al_2010, EtaCar:Skilton12}. 
While the apparent orbital variability effectively rules out the possibility that all emission from \etacar\ is due to the blast wave, a significant fraction of the steady emission could still be produced near the blast wave. Though we do not consider them further in this work, these particles would also contribute to the $\gamma$-ray emission on larger scales.

The paper is structured as follows: In \autoref{sec:env}, we present an overview of the environment around \etacar. This is followed by the analysis of the \fermi\ data in \autoref{sec:fermi}. Afterward, we explore possible $\gamma$-ray emission from escaping particles on scales not resolvable by \fermi\ (\autoref{sec:prop}) and in the molecular clouds on larger scales (\autoref{sec:clouds}). In the end, we discuss our findings in \autoref{sec:discuss}.

\section{Environment around \etacar\ }
\label{sec:env}
\begin{figure*}[h]
    \centering
    
    \includegraphics[width=\textwidth,trim = 20 120 30 80, clip]{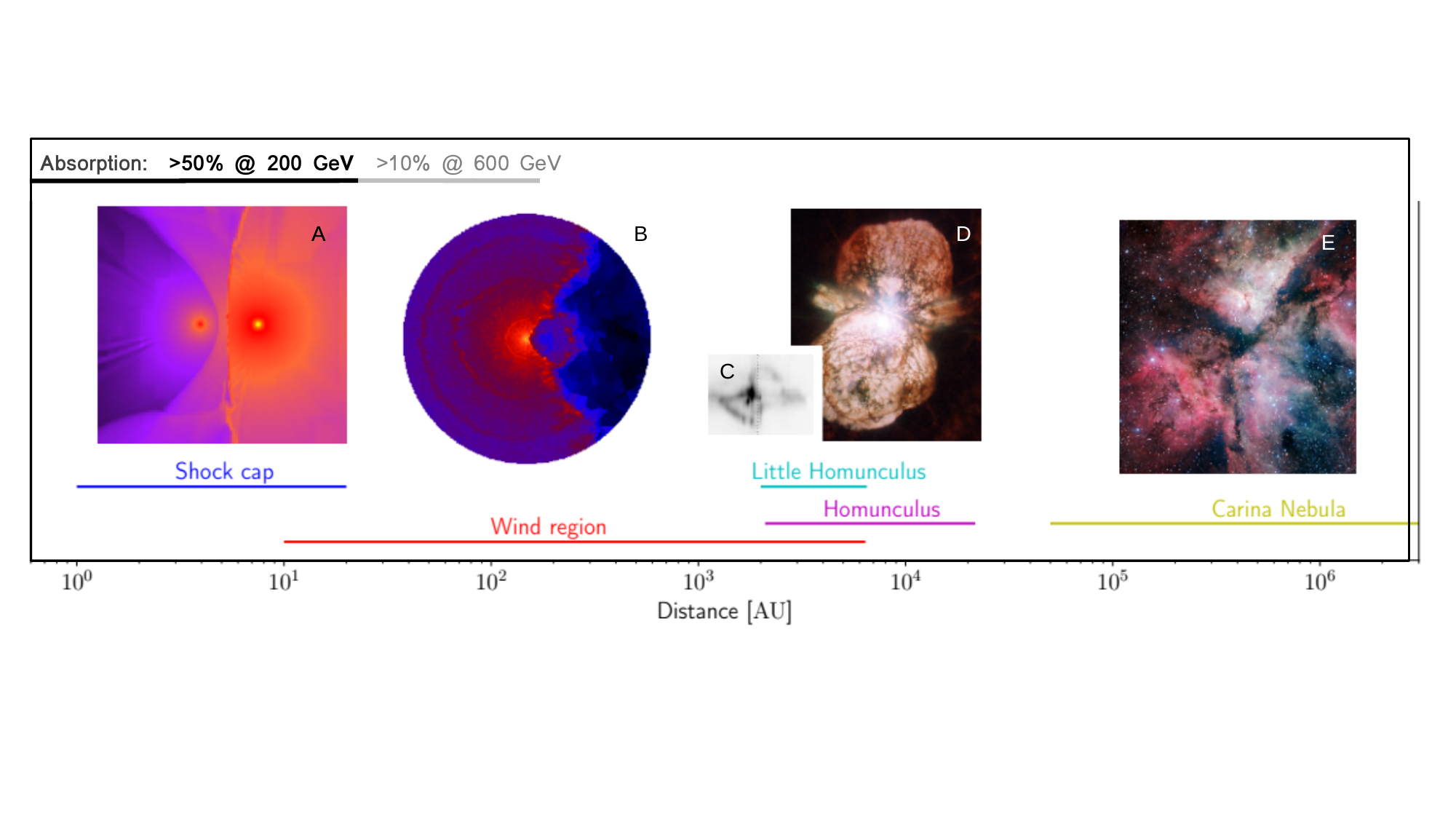}
    \caption{Scales relevant to the possible emission regions around \etacar. Because some regions such as the large or the little Homunculus Nebula are asymmetric, or their size can change over time such as for the shock cap, the sizes shown cover a large range of values. The different extensions are: \num{1} to \SI{20}{\astronomicalunit} (shock cap, A), \num{10} to \SI{6.4e3}{\astronomicalunit} (wind region, B), \num{2e3} to \SI{6.5e3}{\astronomicalunit} (little Homunculus, C), \num{2.1e3} to \SI{21.69e3}{\astronomicalunit} (Homunculus, D) and \SI{> 50e3}{\astronomicalunit} (Carina Nebula, E).  
    The black and grey bars in the upper left corner indicate the distances beyond which the absorption at \SI{200}{\giga\electronvolt} is less than \SI{50}{\percent} (black) and the absorption at \SI{600}{\giga\electronvolt} is less than \SI{10}{\percent} (grey), see text for more details. The individual images are taken from \cite{Parkin_et_al_2011}, \cite{Clementel_et_al_2014}, \cite{Smith_2005_structure_of_little_Homunculus}, HST, ESO. 
    }
    \label{fig:emission_regions}
\end{figure*}

The surroundings in which the \etacar\ binary system is embedded are non-homogeneous with multiple complex structures on a variety of scales. These provide not only a variety of target conditions for CRs escaping from \etacar, but also multiple competing sites for particle acceleration. We consider the following zones, in order of increasing scale, as targets for CR interactions and gamma-ray emission:

\begin{itemize}
    \item {\bf The shock cap/WCR} -- the region where the shocks form between the two stars. Following \RW\ we refer to the more massive star as \etacar-A, and the lower mass companion as \etacar-B. Depicted as region A in \autoref{fig:emission_regions}.
    \item {\bf The pinwheel/wind region} -- a high-velocity outflow with gradual mixing of the high and low-density winds of the two stars, ending on the scale of the little Homunculus (a possible remnant of an outburst more recent than the great eruption). Depicted as region B in \autoref{fig:emission_regions}.
    \item {\bf The Homunculus} -- the expanding shell associated with the great eruption of 1843. Depicted as region D in \autoref{fig:emission_regions}.
    \item {\bf The Carina Nebula / Gum~31 complex} -- the star formation region containing several massive stellar clusters and $10^3$--$10^5$ solar mass molecular clouds  \cite[see e.g.][]{Rebolledo2015}. Depicted as region E in \autoref{fig:emission_regions}.
\end{itemize}

\autoref{fig:emission_regions} illustrates these different regions graphically. 
The approximate sizes of these regions are indicated by the coloured lines below each graphic. Additionally, we illustrate the approximate spatial scales associated with the absorption of the $\gamma$-rays. To compute these scales, the stellar parameters were taken to be the same as in \RW, but additionally the angular size of the stars was taken into account. While the absorption in \RW\ was calculated specifically for the emitting region and integrated for specific orbital phase ranges, here we adopted a spherical geometry of the $\gamma$-ray emission region for simplicity and the two stars were taken to be at their maximum separation, the apastron position. The black line at the uppermost part of the figure shows the radii for which more than \SI{50}{\percent} of $\gamma$-rays with an energy of \SI{200}{\giga\electronvolt} would be absorbed before reaching Earth. The peak absorption occurs close to \SI{600}{\giga\electronvolt}, and the grey line shows the radial distances inside which more than \SI{10}{\percent} of these \SI{600}{\giga\electronvolt} $\gamma$-rays are absorbed. Therefore, accounting for $\gamma$-ray absorption is essential for the emission within \SI{\sim 200}{\astronomicalunit} of the stars. A detailed plot showing the radial dependence for different energies can be found in \autoref{sec:app-b}.

The {\bf shock cap} or {\bf WCR} is formed by the colliding stellar winds. As discussed previously, the winds terminate in a pair of shocks separated by a contact discontinuity. The shock \emph{cap} refers to its finite lateral extent. The global size of this region varies with the orbital phase between \SI{\sim 1}{\astronomicalunit} at periastron and \SI{\sim 20}{\astronomicalunit} at apastron. At the shocks, particles are expected to be accelerated \citep[see e.g.][]{Eichler1993}. A detailed discussion of the particle acceleration and emission can be found in \SO\ and \RW. 

The shock cap and associated pinwheel are embedded in the {\bf wind region}, which is created by the outflowing winds from both stars. Due to the orbital motion, the winds create a pin-wheel like structure, though for \etacare this is complicated by its eccentric orbit \cite[see][]{Parkin2008}. Simulations by \cite{Madura2013} (see also \cite{Parkin_et_al_2011,Clementel_et_al_2014}) find that the fast wind from \etacar-B carves out a low-density channel with a large opening angle in the direction of the apastron position. This is a consequence of the brief periastron passage. As is evident in image B in \autoref{fig:emission_regions}, the darker region indicates the low-density wind from \etacar-B. Accelerated particles that enter this region experience negligible collisions. On the other hand, any particles that encounter denser wind material from \etacar-A within $\approx$100~au from the binary will be collision dominated and have a low probability of escaping the wind region.

The wind region is expected to end as it reaches the {\bf little Homunculus Nebula} (see  \autoref{fig:emission_regions} C). This nebula has a total mass of plausibly \SI{\sim 0.1}{\solarmass} \citep{Smith_2005_structure_of_little_Homunculus}. It is believed to be ejected in the outburst from 1890 and accelerated by the stellar winds. It has a bipolar shape, with the equator being \SI{\sim 2e3}{\astronomicalunit} and the poles around \SI{6.5e3}{\astronomicalunit} away from the central stars.

The little Homunculus is in turn embedded in the larger {\bf Homunculus Nebula}, the large bipolar structure shown in  image D of \autoref{fig:emission_regions}. The Homunculus was ejected during the great eruption of \etacar\ in the 1840s. It is mainly composed of a thin outer shell, with a thickness of up to \SI{600}{\astronomicalunit}, and a thicker inner shell \citep{Smith_2006}. The thin outer shell contains \SI{\sim 90}{\percent} of the total mass of the Homunculus. The total mass is not exactly known, but it is larger than \SI{10}{\solarmass} \citep{EtaCar:Smith03} and thought to be between \SI{15}{\solarmass} and \SI{35}{\solarmass} \citep{Smith_Ferland_2007}. This makes the Homunculus the most massive structure in the close proximity of the stars. The equivalent pure atomic hydrogen densities in the thin outer shell are high: at least \SI{3e6}{\per\cubic\centi\meter} but potentially \SI{e7}{\per\cubic\centi\meter} or even higher \citep{Smith_2006}.

The Homunculus lies within the {\bf Carina Nebula}. As detailed in \cite{Rebolledo2015} (section 5.3), four nearby clouds with masses of $\approx$10$^4$-$10^5 \, \mathrm{M_\odot}$ are detected in CO maps and also dust measurements. These are referred to as the Southern Cloud, Northern Cloud, Southern Pillars, and Gum\,31 and are located within 100~pc of \etacar. From the velocity maps presented in that paper, a clear connection to the CNC/Gum\,31 complex is evident. Whereas the emission of the binary system down to the scale of the Homunculus is not resolvable with Fermi or \hess, the larger clouds allow us to probe the CR density around \etacar\ as discussed in \autoref{sec:fermi} and \autoref{sec:clouds}.

\section{Fermi analysis and results}
\label{sec:fermi}

The approach to Fermi data analysis here follows closely that of \RW. The data selection was based on the latest \fermi\ Pass 8 data \citep{Bruel2018} starting from Aug 4$^{\rm th}$, 2008 (MET 239557417) to Oct 26$^{\rm th}$, 2022 (MET 688521600). Events over an energy range of 500 MeV (chosen to avoid poorly reconstructed events at lower energies) and 500 GeV were included from a region of interest (ROI) of \SI{10}{\degree} by \SI{10}{\degree}, centred at the nominal position of \etacar\ and aligned in galactic coordinates. Data were chosen according to the SOURCE event class ({\it evclass=128}) with FRONT+BACK event types ({\it evtype=3}). Time periods in which the ROI was observed at zenith angle greater than \SI{90}{\degree} were excluded. Data were analysed using {\it Fermitools} (version 2.2.0), which is the official {\it ScienceTools} suite provided by the Fermi Science Support Center\footnote{\url{https://fermi.gsfc.nasa.gov/ssc/}} and {\it FermiPy} (version 1.2) \citep{Fermipy2017}. The model of sources surrounding \etacar\ was taken similar to  \RW\ and includes all sources in the ROI from the \emph{Fermi}-LAT 12-year source catalog (4FGL-DR3) \citep{4fgldr3}, with the exception of the unidentified point source 4FGL J1046.7-6010, which is located in the centre of the Southern Pillars and was excluded to avoid a potential miss-assocation of diffuse flux in this region. 
The model including the diffuse Southern Pillars component instead of the point source 4FGL J1046.7-6010 is preferred by $\Delta TS = 27.3$, whereas adding the point source additionally only improves the model by $\Delta TS = 7.0$ with 3 additional degrees of freedom. More details on the analysis configuration are summarized in \autoref{sec:app-a}. 

\begin{figure}
    \centering
    \includegraphics[width = 0.5 \textwidth]{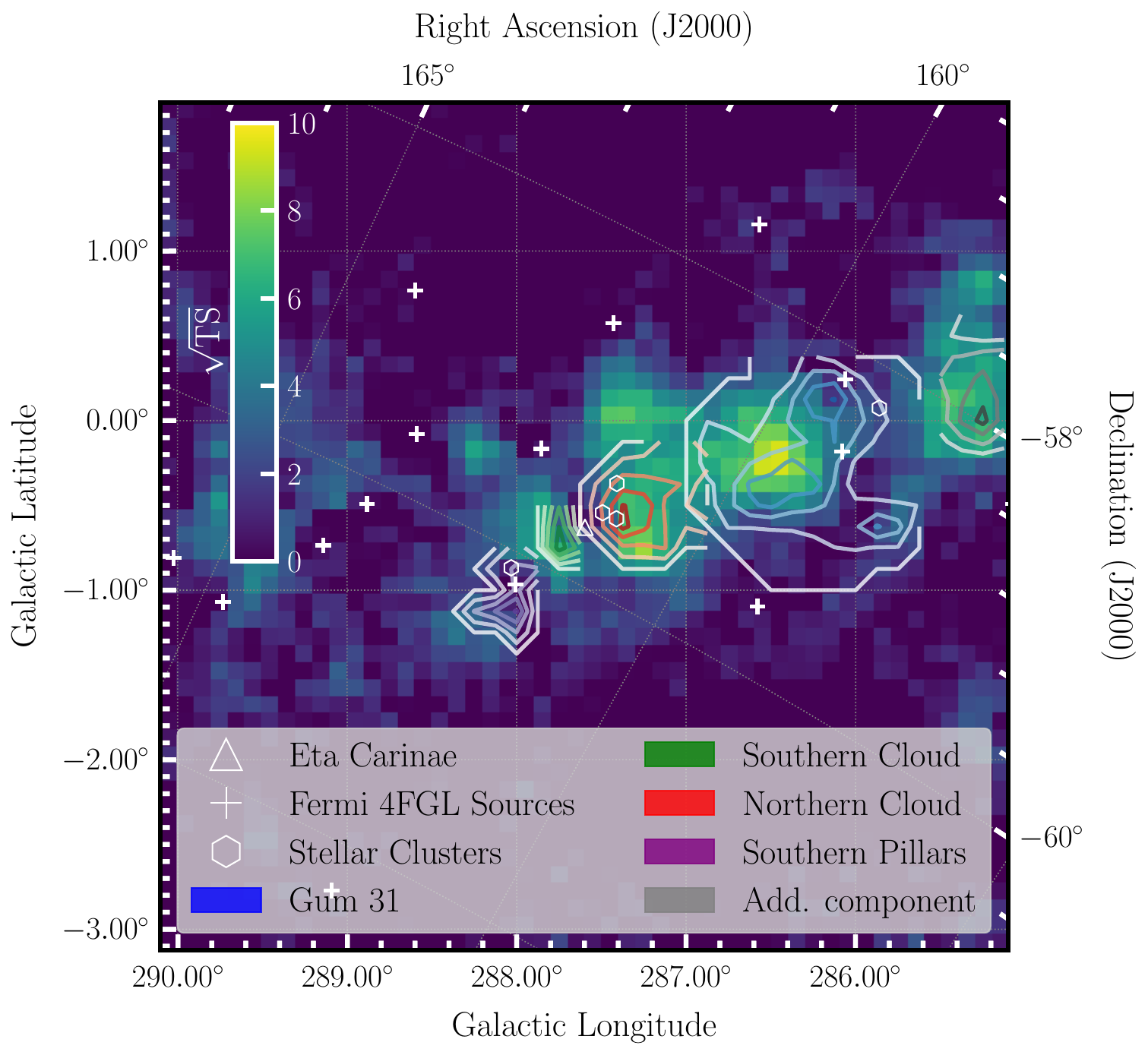}
    \includegraphics[width = 0.5 \textwidth]{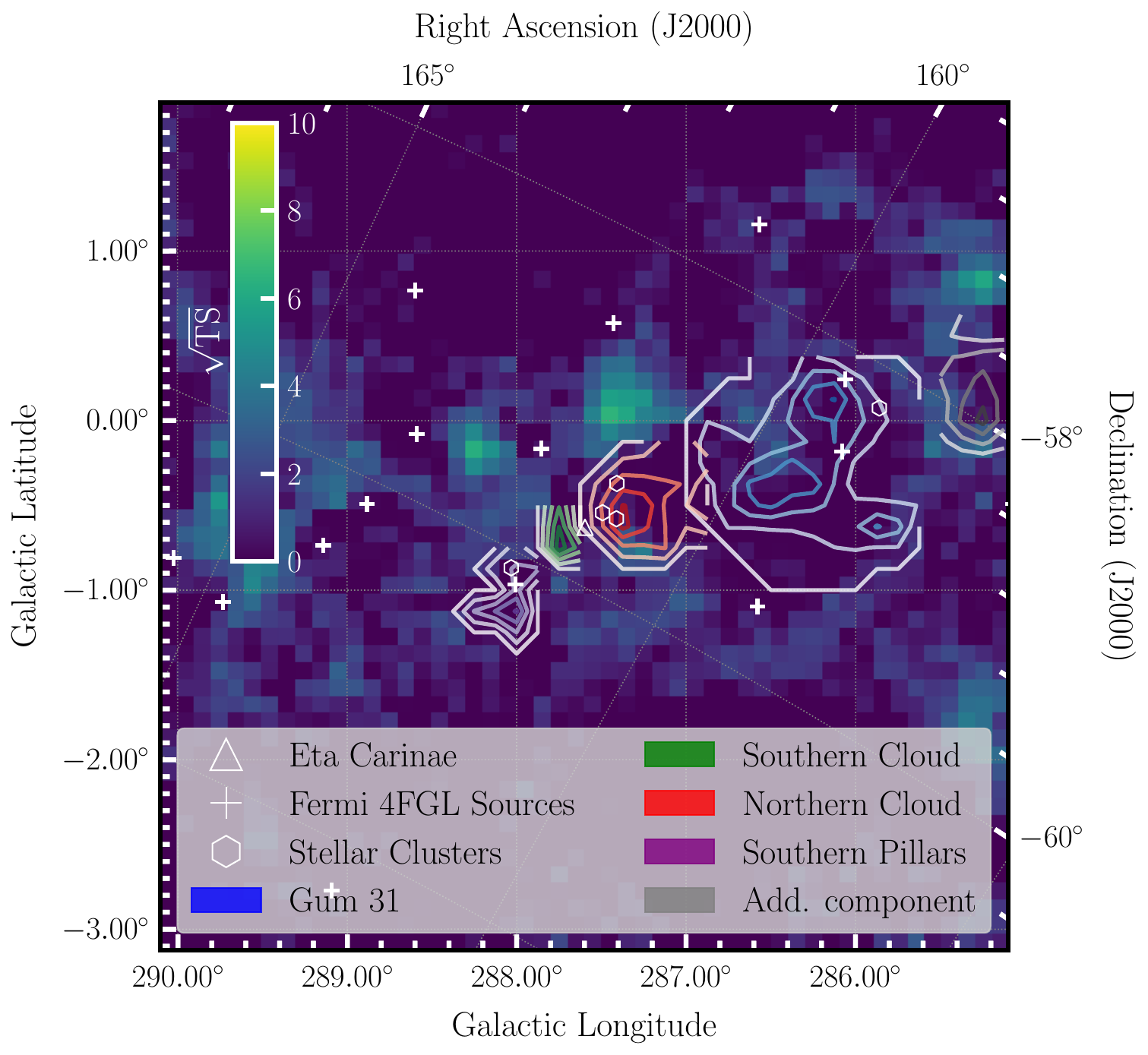}
    \caption{Residual significance maps from the \fermi\ analysis in a zoomed region of the ROI. In the upper panel, the $\sqrt{\rm TS}$ map without the addition of the cloud templates to the model is shown, whereas they are included in the lower panel. In each plot the cloud templates are overlaid by coloured contours, indicating the gas density distribution. Additionally, all other 4FGL sources 
    are marked together with massive stellar clusters from \cite{Preibisch2011}.}
    \label{fig:fermi_res}
\end{figure}

In \RW\ an additional diffuse source was added to the model based on the CO survey of \citet{Dame2001}. Here, we split the model into four individual clouds, Southern Cloud, Northern Cloud, Southern Pillars and Gum 31, following the region definitions as outlined in \cite{Rebolledo2015} and described in the previous section. The four templates can be seen as an overlay in \autoref{fig:fermi_res}. Each cloud is included in the model as a diffuse component with a power-law spectral shape.
Optimisation of the model was performed in an iterative fashion making use of the {\it optimize} function of {\it FermiPy}. First, the normalisation of a maximum of 5 of the brightest sources with a predicted number of counts totaling 95\% of the total predicted counts of the model are freed and a simultaneous fit is performed. Next, the normalisation of the sources  not included in the first step was performed individually. Finally, the shape and normalisation of all sources with a TS exceeding 25 in the previous fits were freed and a simultaneous fit was performed. The same process was repeated but allowing up to ten sources for the first optimization step. After optimising the model, the cloud component spectral shapes were fixed in a first fitting iteration freeing \etacar\ (4FGL J1045.1-5940) and the normalisation of all sources within a 3$^{\circ}$ radius and with TS values $>$10 and more than 10 predicted counts. As in the 4FGL catalogue, and previous analysis of the region  \cite[\RW, ][]{MartiDevesa21}, \etacar\ was modelled as a point source with a log-parabola spectrum $\phi=\phi_0(E/E_0)^{-\alpha-\beta\log(E/E_0)}$. This yielded a best fit model for \etacar\ with $\alpha = 2.29 \pm 0.02$, $\beta = 0.11 \pm 0.01$, $\phi_0 = (2.44 \pm 0.07) \times 10^{-6}  \, \UNITS{cm^{-2}\,erg^{-1}\,s^{-1}}$  and $E_0 = 2.11 \, \UNITS{GeV}$ consistent with previous studies \cite[\RW, ][]{MartiDevesa21}. In a second fitting iteration \etacar\ and all other previously free sources were fixed and the normalisation and spectral shape of all four clouds were freed. Hence any miss-association of flux between the close Southern and Northern Clouds and the bright \etacar\ could be minimised. Following the fitting for \etacar\ and the molecular clouds, a spectral energy distribution (SED) was generated for the source of interest. 
Whereas a similar event selection approach was taken in \citet{Ge_et_al_2022_clouds_around_EtaCar}, their main result is based on the addition of two Gaussian disks in the fit instead of the cloud templates employed in this work. 

\begin{figure}
    \centering
    \includegraphics[width = 0.5 \textwidth]{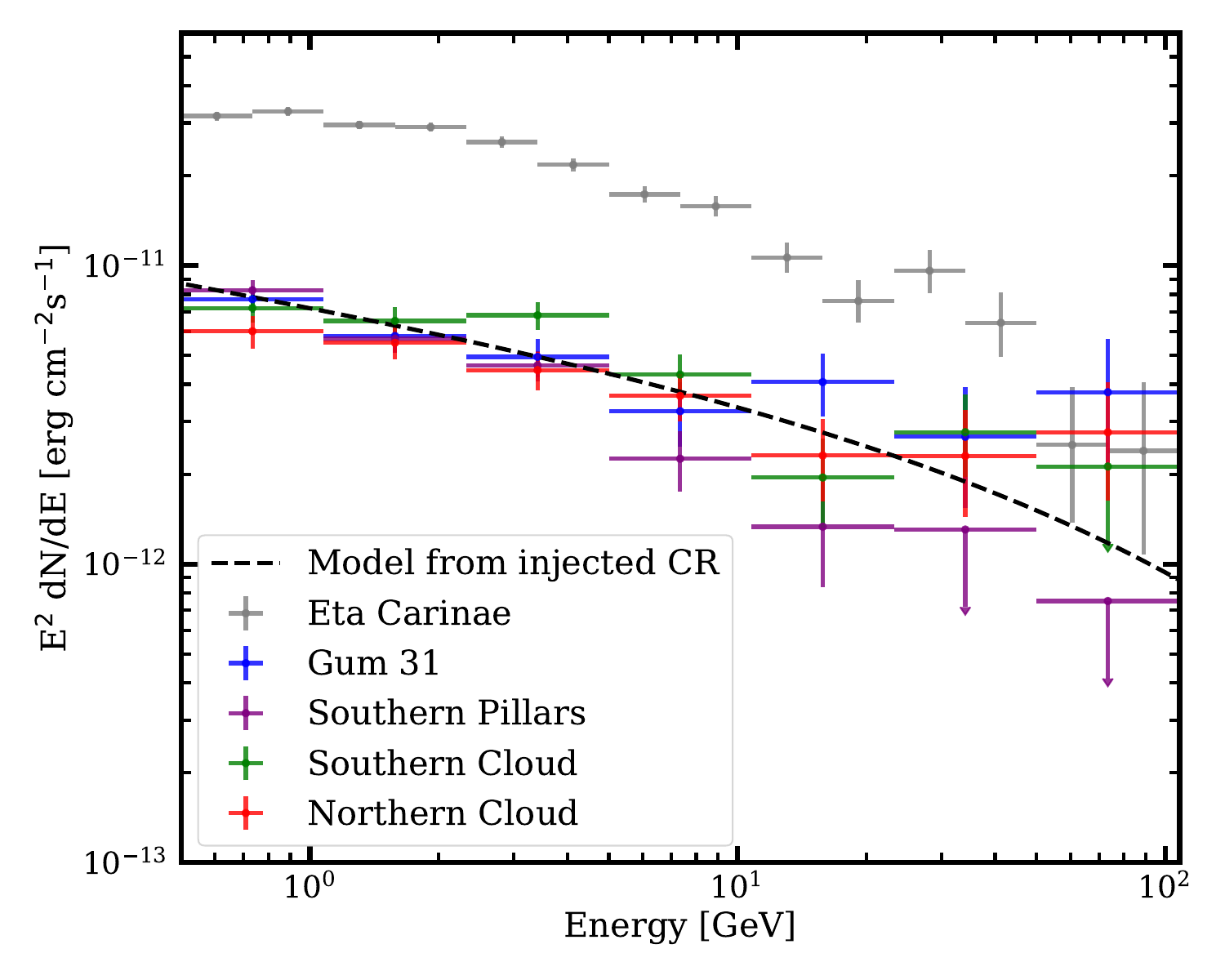}
    \caption{Spectra of the four molecular clouds as derived from the \fermi\ analysis.
    For comparison, the spectrum derived for \etacar\ is also shown in grey. Additionally, a model derived from an injected CR spectrum is shown. The CR spectrum is modeled as a power law with an index of 2.0 and an exponential cutoff at 2~TeV. The normalization has been scaled arbitrarily.  }
    \label{fig:fermi_flux}
\end{figure}

The four clouds were each detected with significant excesses and TS values of $>$200. The exact values are summarised in \autoref{tab:cloud_spectra}. If a photon weighting procedure as used in the 4FGL catalog \citep{4FGL2020} is applied the TS values are reduced but still $>$150 for each cloud.
The residual emission in the ROI could be clearly reduced with the addition of the cloud templates, as shown in \autoref{fig:fermi_res}. 
The remaining residual emission to the right side of the region can not be associated with molecular clouds in the  CNC/Gum 31 complex \citep{Rebolledo2015} but might be connected to diffuse emission around Westerlund 2 \citep{Westerlund2}. In order to account for it in the fit an additional diffuse component based on the CO map by \citet{Dame2001} was introduced. The inclusion of this additional component helped to improve the overall fit but did not affect the properties of the cloud components significantly.
The spectra of the four individual clouds are shown in  \autoref{fig:fermi_flux} and the best-fit spectral parameters are also included in \autoref{tab:cloud_spectra}. The spectral results with and without the photon weighting differ only by up to 0.04 for the spectral index and \SI{3}{\percent} on the energy flux, hence being consistent within the statistical uncertainties. The values without the additional weighting are adopted, whereas the quoted differences can be interpreted as a contribution to the systematic uncertainty. For comparison, the spectrum of \etacar\ is also shown. The spectral shapes and normalisations determined for the Southern Cloud, Northern Cloud and Gum\,31 regions show a reasonable level of consistency, with a spectral index between 2.2 and 2.3. In contrast, the Southern Pillars are described by a slightly softer spectrum, mostly caused by an apparent cutoff above photon energies of 20 GeV. Below that energy, the derived flux points are comparable amongst all four clouds.  
Different spectral models for each cloud were tested by fitting the SEDs of each cloud separately. This yielded a difference of as much as $~10 \%$ for the integrated energy flux, which can be considered as a contribution to the systematic uncertainty of the analysis. 

\begin{table}
    \centering
    \begin{tabular}{c|c|c|c}
        Cloud Name & Spectral & Energy Flux & TS \\
        & Index & [$\UNITS{erg\,cm^{-2}\,s^{-1}}$] \\
        \hline 
        Southern Cloud & $2.30 \pm 0.05$ & $(2.4 \pm 0.3) \times 10^{-11}$ & 276 \\
        Northern Cloud & $2.25 \pm 0.06$ & $(2.0 \pm 0.2) \times 10^{-11}$ & 225\\
        Southern Pillars& $2.54 \pm 0.06$ & $(1.8 \pm 0.2) \times 10^{-11}$ & 348\\
        Gum 31 & $2.24 \pm 0.06$ & $(2.4 \pm 0.3) \times 10^{-11}$ & 270\\
    \end{tabular}
    \caption{TS values and power-law spectral properties of the cloud templates as included in the model and derived by the fit. The energy flux is derived by integrating from 500 MeV to 100 GeV. }
    \label{tab:cloud_spectra}
\end{table}

The inner regions of the Carina Nebula are not resolvable with current $\gamma$-ray instruments. For example, the Homunculus exists on scales corresponding to a factor 40 smaller than the 68\,\% containment radius of the Fermi Point Spread Function (PSF), which is  ${\sim} 0.1^{\circ}$ above 20~GeV \citep{4FGL2020}.
An upper limit on the $\gamma$-ray flux originating from this inner region (but outside the WCR) can hence only be estimated via the quiescent baseline component of the \etacar\ $\gamma$-ray flux. Long-term light curves show only mild variability \cite[e.g. \RW,][]{MartiDevesa21}, with the $\gamma$-ray flux never dropping below $~60 \%$ of the mean integrated flux level. This provides a baseline energy flux of $(5.6 \pm 0.3) \times 10^{-11} \UNITS{erg\,cm^{-2}\,s^{-1}}$ integrated from 500~MeV to 100~GeV.

\section{Propagation of CRs around $\eta$ Car and expected emission}
\label{sec:prop}

\subsection{Particle acceleration and emission in the WCR}
\label{sec:prop_wcr}
We summarise in the following the relevant points of the model of \SO, \RW\ for CR production in the WCR \cite[see also][]{Eichler1993,Bednarek11}. The model from \RW\ is then used to obtain the spectrum of escaping CRs.

As mentioned in \autoref{sec:env}, the WCR contains two shocks. The wind of the more massive star \etacar-A is slower (terminal wind speed $v_{\infty} \approx \SI{5e2}{\kilo\meter\per\second}$) and more dense ($\dot M \approx \SI{4.8e-4}{\solarmass\per\year}$) compared to that of the companion star \etacar-B  \cite[$v_{\infty} \approx \SI{3e3}{\kilo\meter\per\second}$, $\dot M \approx \SI{1.4e-5}{\solarmass\per\year}$, see][]{Parkin_et_al_2011}. At the side of \etacar-A, the high density leads to complete cooling of accelerated protons. The maximum proton energy is limited by inelastic collisions with compressed wind material, with $E_{\rm max} \sim 200$~GeV, and no energetic particles escape the acceleration region. On the side of \etacar-B, the cooling time via  collisions greatly exceeds the flow time, and accelerated particles can leave the acceleration region. \RW\ estimated that the maximum proton energy achievable at the wind termination shock of \etacar-B is $\approx 30$\,TeV.

To match observations, approximately \SI{10}{\percent} of the wind power processed by each of the shocks should be converted into accelerated protons above $1$\,GeV. 
While energetic particles on the side of \etacar-A are completely destroyed via hadronic collisions, on the side of \etacar-B  particles can enter the ballistic region, i.e. the flow transporting the shocked material laterally away from the WCR.
The accelerated protons from the shock of \etacar-B produce a non-negligible amount of $\gamma$-rays in the ballistic region as the shocked material from \etacar-B mixes with the shocked higher density material from \etacar-A. The amount of radiation produced depends on the details of the mixing process, which was adjusted in \RW\ to match the \fermi\ data. The number of interacting particles in the \RW\ model varies with phase and is largest at periastron.

For $\gamma$-rays produced in the wind region close to the stars, absorption due to pair production plays an important role. Here, we restrict ourselves to the assumption of spherically symmetric emission.
Due to the change in the stellar positions, there is an orbital variation in the absorption - in particular between phases 0.9 and 0.2. However, for emission on radial distance scales of \SI{20}{\astronomicalunit}, the phase dependence drops below \SI{7}{\percent} and decreases rapidly for emission on even larger scales. 
 
In \autoref{fig:emission_regions} the absorption for spherically symmetric emission at phase \num{0.5} for two different energies is shown. The case of \SI{600}{\giga\electronvolt} corresponds roughly to maximum absorption. Whereas at \SI{20}{\astronomicalunit} close to half of the emission is absorbed, at \SI{200}{\astronomicalunit} more than \SI{90}{\percent} of the $\gamma$-rays reach Earth.
Thus, the VHE $\gamma$-ray emission reported by \citet{HESS_2020_EtaCar} can be easier accounted for if it is produced at larger radii.

In the following, we use the average escaping CR spectrum and the stellar wind parameters quoted above, adapted from the model of \RW, to investigate possible $\gamma$-ray emission produced by the escaping protons in the surrounding environment. 
Because the escaping proton spectrum in that work was not calculated directly, here it was inferred from the $\gamma$-ray data presented in \RW. To obtain the spectral shape, we used the Naima package \citep{naima} assuming an exponential cutoff power-law proton spectrum. In the off-periastron phase range, this resulted in a power-law index of $\alpha\approx 2$ and cutoff energy of $E_{\rm cut}\SI{\approx 1.8}{\tera\electronvolt}$. In \RW\ the $\gamma$-ray spectrum is calculated while the particles are advected with the flow out to a distance where the resulting emission is negligible and the radiative energy losses do not further change the underlying spectrum. The spectrum of escaping particles is then assumed to maintain the same shape. The proton spectrum above \SI{1}{\giga\electronvolt} was then normalised such that the total energy in escaping protons and transferred into $\gamma$ rays, secondary particles and directly accelerated electrons is equal to the total power put into particle acceleration. Approximately \SI{\sim 90}{\percent} of the accelerated protons at the side of \etacar\ B escape from the WCR.

\subsection{Propagation in the wind region}

The energetic particles that escape in the ballistic flow will eventually merge with the winds. To avoid substantial adiabatic losses, the transport of these particles should be diffusion dominated. The ratio of the characteristic advection and diffusion timescales is a useful measure of which process dominates the radial transport:
\begin{align}
\frac{t_{\rm adv}}{t_{\rm diff}} = 
\frac{R}{u} \times \frac{D_r}{R^2} = \frac{D_r}{uR}
\end{align}
where $u$ is the outward flow speed, $D_r$ the average radial diffusion coefficient, and $R$ the radial distance from the binary. Taking $u \sim 1000\,{\rm km\, s}^{-1}$ and $R=100$\,au, adiabatic losses can be neglected provided $D_r \gg 10^{23} {\rm cm^2\, s^{-1}}$. While the radial diffusion coefficient is not known, the magnetic fields will be highly disordered in the ballistic region provided mixing is effective. This can in principle allow relativistic protons and other nuclei to decouple from the flow and escape the wind region without undergoing significant adiabatic losses.

If, on the other hand, particles were strongly coupled to the flow, the maximum energy would be reduced by adiabatic cooling: $E\propto \rho^{1/3}\propto r^{-2/3}$. Thus in crossing the wind region, the maximum particle energy would be reduced by a factor of 10 or more. If the VHE measurements of \citet{HESS_2020_EtaCar} are produced far from the WCR, this provides a (model-dependent) constraint on the transport.

Protons that move outward, by either diffusion or advection, but remain in the low-density cavity evacuated by the wind from \etacar-B, will produce negligible $\gamma$-ray emission. 
However, some protons might migrate into the high-density wind from \etacar-A, radiate significantly, and if they remain in the high-density zone, lose all of their energy via hadronic interactions. 

The spectrum of the protons migrating into the \etacar-A wind will depend on the transport processes, the details of which are currently not well-constrained. In the most extreme case, when all protons penetrate into the wind of \etacar-A, the energy in $\gamma$ rays detected on Earth using a total energy flux in escaping particles of \SI{6.5e35}{\erg\per\second} from the model of \RW\ would be \SI{e-9}{\erg\per\square\centi\meter\per\second}. This exceeds considerably the flux detected by \fermi , and the model parameters of \RW\ would have to be adapted. In the other extreme case, if a transport barrier exists at the interface of the two winds, excluding CRs from the denser target material, no significant emission would be produced in the wind region.

Variability of the $\gamma$-ray emission in the wind region are likely to be modest and not necessarily linked to the orbital phase, unless the $\gamma$ rays are produced only very close to the stars. Any potential emission from escaping protons in the wind region will thus contribute to an approximately steady baseline flux, modulo turbulent variations. The same should hold true for larger structures such as the little and large Homunculus nebulae.

\subsection{The little and large Homunculus nebulae}
As shown in the previous section, only the particles diffusing in the low-density wind from \etacar-B will leave the pinwheel/wind region and in the most extreme case, this could be the vast majority of the particles leaving the WCR.
After the wind region, the CRs encounter the little Homunculus Nebula, which is likely to have a total mass of \SI{\sim 0.1}{\solarmass} \citep{Smith_2005_structure_of_little_Homunculus}. 
On larger scales they encounter the $>$\SI{10}{\solarmass} Homunculus Nebula
(see \autoref{sec:env}). 
To obtain a rough estimate of the total $\gamma$-ray emission in each region, we can use the following approximation:
\begin{align}
    F_{\rm Earth} \approx \frac{t_{\rm cross}}{t_{\rm cool}} \cdot \frac{L_{\rm esc}}{4\pi d^2},
\end{align}
where $F_{\rm Earth}$ is the total energy flux on Earth, $t_{\rm cross}$ the time it takes for the particles to cross each region, $t_{\rm cool}$ the cooling time, $L_{\rm esc}$ the total energy in escaping particles and $d = \SI{2.3}{\kilo\pc}$ the distance towards \etacar. For $L_{\rm esc}$ we adopt the value of $\SI{6.5e35}{\erg\per\second}$ from the model of \RW, assuming that energy losses are negligible in the pinwheel/wind region.
We model the little Homunculus and the inner and outer shells of the large Homunculus as spherical shells around the stars for simplicity. We first consider the case of no diffusion within these thin shells, with CRs simply passing through at (close to) the speed of light.

\cite{Smith_2005_structure_of_little_Homunculus} estimated a thickness of the little Homunculus of \SI{\sim 940}{\astronomicalunit}. Using an average inner radius of \SI{4e3}{\astronomicalunit}, a thickness of the shell of \SI{1000}{\astronomicalunit} and a mass of \SI{0.1}{\solarmass}, the cooling time at \SI{1}{\giga\electronvolt} is \SI{\sim 4e5}{\day}. If the CRs move through the region with the speed of light, the resulting total energy flux on Earth is $F_{\rm Earth} = \SI{1.4e-14}{\erg\per\square\centi\meter\per\second}$. This is orders of magnitudes below the energy flux from \fermi\ of \SI{5.6e-11}{\erg\per\square\centi\meter\per\second}.
The inner shell of the large Homunculus contains \SI{10}{\percent} of the total Homunculus mass \citep{Smith_2006}, which is between \SI{1}{\solarmass} and \SI{3.4}{\solarmass}. To model it as a shell, we assume an inner radius of \SI{1e4}{\astronomicalunit} and an outer radius of \SI{1.5e4}{\astronomicalunit}. For a mass of \SI{\sim 2}{\solarmass}, the resulting $F_{\rm Earth}$ is \SI{3.7e-14}{\erg\per\square\centi\meter\per\second}. Different values of the inner and outer radius of the shell give similar results. For the outer Homunculus shell with an assumed density of \SI{1.0e7}{\per\cubic\centi\meter} and a thickness of \SI{600}{\astronomicalunit}, one obtains $F_{\rm Earth} = \SI{6.2e-13}{\erg\per\square\centi\meter\per\second}$. The outer Homunculus shell is therefore the most important region for the production of $\gamma$ rays by the escaping particles with fluxes at least one order of magnitude above the fluxes from the little Homunculus and the inner Homunculus shell.
However, if the CRs pass through the outer shell at the speed of light, the $\gamma$-ray emission produced is still two orders of magnitude below the \fermi\ flux.

The Weigelt knots, located closer to the binary, have not been explicitly considered in this study. Adopting the typical Weigelt knots parameters from \citet{Remmen13}, the knots do not appear to have a sufficiently high over-density relative to the parameters adopted for the Homunculus Nebula to make a significant difference to the gamma-ray luminosity on those scales.

 Cosmic rays are more likely to diffuse slowly through these shells of material, than pass through at the speed of light. The outer Homunculus shell is the most massive and the most dense of the three shells, with likely larger magnetic fields and slower diffusion and hence expected to dominate also in this case.
 We therefore restrict the following calculations for different diffusion coefficients to emission from the outer shell. For simplicity, we ignore possible spectral changes caused by the propagation through the wind region, the little Homunculus and the inner Homunculus shell, because the exact propagation properties are unknown and are impossible to separate observationally from spectral changes by the propagation through the Homunculus itself. 
As before we adopt a value of \SI{1.0e7}{\per\cubic\centi\meter} for the density in the outer shell. 
A lower limit on the (parallel) diffusion coefficient can be estimated by assuming Bohm diffusion. Unfortunately, the magnetic field in the shell is unknown and difficult to constrain. Values of \SI{100}{\micro\gauss} or even higher have been proposed \citep{Aitken_et_al_1995}. To provide some numerical estimates, we take the energy dependence of the radial diffusion coefficient to have the following form:
\begin{align}
\label{eq:diff}
    D_r(E) = D_0 \times \left(\frac{E}{\SI{1}{\giga\electronvolt}}\right)^{\alpha}.
\end{align}

\begin{figure}
    \centering
    \includegraphics[width=0.5\textwidth]{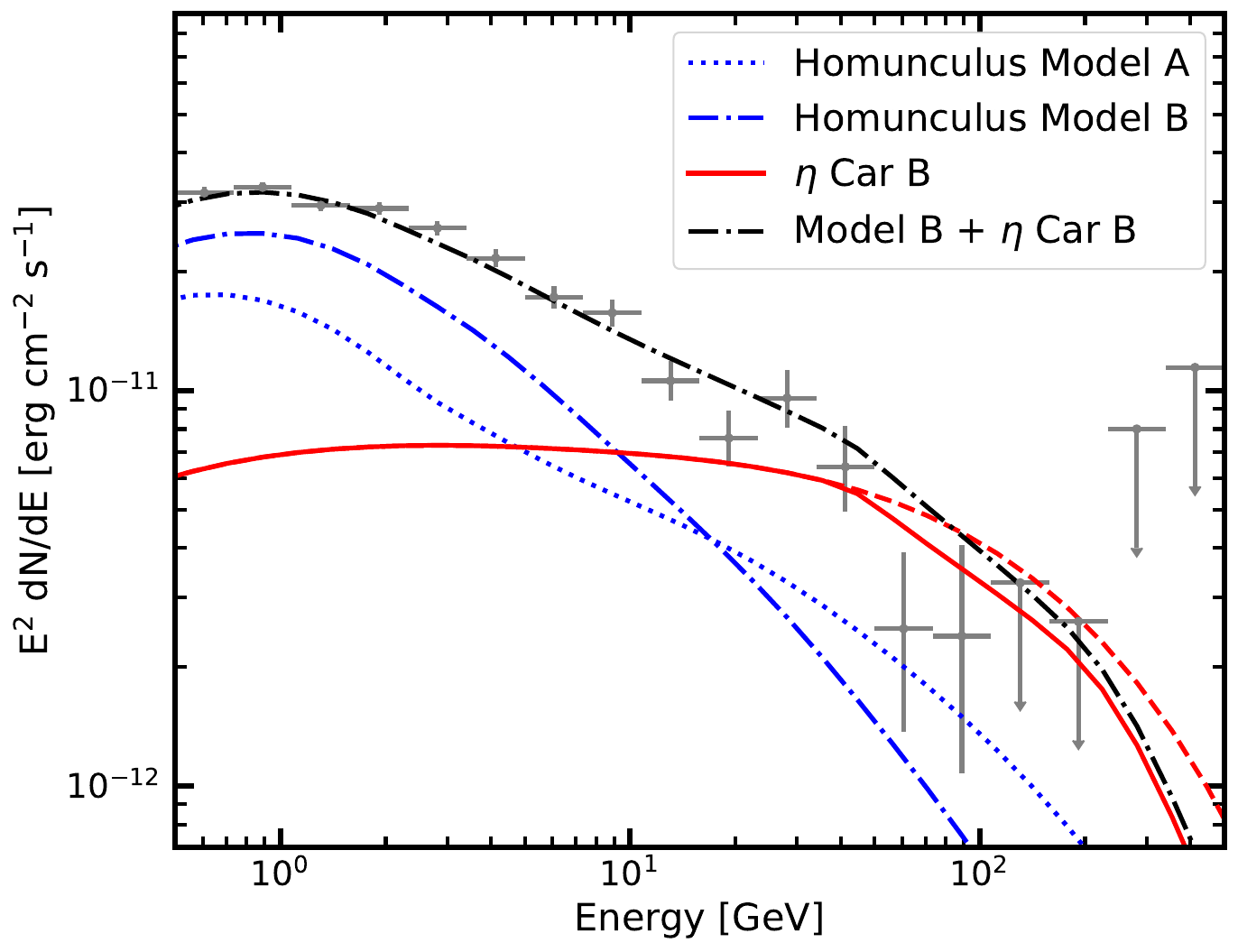}
    \caption{Model of emission from \etacar\ and the Homunculus together with \fermi\ data. The blue curves show emission from the Homunculus for different diffusion properties of the escaping CRs. The dotted curve (Model A) is for $D_0 = \SI{9e22}{\square\centi\meter\per\second}$ and $\alpha = 0.5$. The dashed-dotted curve (Model B) is for $D_0 = \SI{9e21}{\square\centi\meter\per\second}$, $\alpha = 1$ and a factor of \num{1.1} more escaping CRs compared to the model from \RW. The red solid line shows the emission from \etacar\ produced by particles accelerated at the shock towards \etacar\ B from \RW\ (the red dashed line is without absorption). The dashed-dotted black curve shows the combined emission from \etacar\ B and Model B.}
    \label{fig:homunculus_emission}
\end{figure}

The dotted blue curve in \autoref{fig:homunculus_emission} shows the emission produced in the Homunculus for $D_0 = \SI{9e22}{\square\centi\meter\per\second}$ and $\alpha = 0.5$. As can be seen, interactions in the Homunculus Nebula of escaping CRs could produce a significant amount of steady $\gamma$-ray emission. Thus, models in which particle acceleration at the shock of \etacar-A is ineffective, for example, due to a low Mach number or a highly oblique magnetic field, can be replaced with models in which the lower energy $\gamma$-rays are principally produced in the Homunculus. In \autoref{fig:homunculus_emission} the red curve shows the emission produced by particles accelerated at the shock of \etacar-B from the model of \RW, the dashed red line is without absorption. The black dashed-dotted line shows this emission added to the emission from the Homunculus shown in the blue dashed-dotted curve for $D_0 = \SI{9e21}{\square\centi\meter\per\second}$, $\alpha = 1$ and a factor of \num{1.1} more escaping CRs compared to the model of \RW. The combined emission reproduces well the \fermi\ data over the whole energy range. In such a scenario, the weak phase-dependent variability is entirely produced by changes in the emission from \etacar-B close to the binary. The assumed diffusion coefficient in the Homunculus is well above the case of Bohm diffusion for a magnetic field of \SI{100}{\micro\gauss}.

The above model for particle acceleration occurring solely at the companion shock together with steady emission from the Homunculus by the escaping CRs can also account for the hard non-thermal X-ray emission. In the model of \RW\ the hard X-rays, similar to the $\gamma$-rays, are dominated by particles accelerated at the shock of \etacar-B. Since that model reproduced the variability, we focus here on the phase averaged emission \cite[see also][]{BreuhausThesis}. While the orbital variability in the $\gamma$-ray emission shows energy-dependent behaviour that differs between orbits  \citep{Balbo2017,MartiDevesa21}, observations are nevertheless consistent with at least \SI{\sim 60}{\percent} of the emission being produced by a steady-state source. 
Using the existing $\gamma$-ray data, it is difficult to conclude which of these components dominates the quasi-steady emission, e.g. emission produced near the wind termination shock of \etacar-A \cite[\RW, ][]{Balbo2017}, or that produced on larger scales as presented here \cite[see also][for additional large scale sources]{Ohm_et_al_2010, EtaCar:Skilton12}.
An observation of a reduction of the steady part of the emission or a clearly different behavior in different energy bands could potentially constrain the possible contribution of the Homunculus Nebula and the wind region to the gamma-ray flux, though the latter could equally be explained by the turbulent behaviour during the passage as suggested by \citet{MartiDevesa21}.

\section{Propagation in the Carina Nebula}
\label{sec:clouds}

CRs that escape both \etacar\ and the Homunculus Nebula will eventually enter the Carina Nebula. There they will encounter the molecular clouds described in \autoref{sec:env} as potential target material for further $\gamma$-ray production. 
The clouds of the CNC and Gum\,31 exhibit significant $\gamma$-ray emission (see \autoref{sec:fermi}) which could be related to \etacar.
\autoref{fig:fermi_flux} also contains a model $\gamma$-ray spectrum derived from an injected CR distribution following a power-law with exponential cutoff. In the models of \SO\ and \RW\ a non-linear DSA model was used to account for the required hard underlying proton spectrum (see \autoref{sec:prop_wcr}). The assumed CR spectrum has an index of 2.0 and an exponential cutoff at 2~TeV. To compute a $\gamma$-ray spectrum the parametrizations from \cite{Kappes2007} were employed and energy-dependent propagation leading to a softening of $-1/3$ was assumed. 
The resulting $\gamma$-ray spectrum has a similar profile to that measured by \fermi.
The CR density in a certain region can be calculated from the $\gamma$-ray luminosity and cloud mass following the approximation given in \cite{Aharonian2019}:
\begin{equation}
\scriptstyle
    w_{CR}(\geq 10 E_\gamma) = 1.8 \times 10^{-2} \left(\frac{\eta}{1.5}\right)^{-1} \left(\frac{L_\gamma(\geq E_\gamma)}{10^{34} \UNITS{erg\,s^{-1}}}\right) \left(\frac{M}{10^6 M_\odot}\right)^{-1} \UNITS{eV\,cm^{-3}}
\end{equation}
Here the parameter $\eta$ accounts for the presence of heavier nuclei and is assumed to be 1.5. The $\gamma$-ray luminosity, L$_\gamma$, above 500 MeV is derived from integrating the $\gamma$-ray flux and assuming a typical distance of 2.3~kpc, i.e. the same as \etacar. This translates to a minimum CR energy of 5~GeV. The masses are assumed to follow the dust mass estimate from \cite{Rebolledo2015}, Table 2 based on Herschel infrared Galactic Plane Survey (HiGAL, \cite{Molinari10}) data. The mass uncertainty derived from dust maps is mostly dominated by uncertainties in temperature derivation ($\sim$10$\%$) \citep{Uruqhart18} the HiGAL survey flux ($\sim$5$\%$) \citep{Molinari16} and the local gas-to-dust mass ratio assumption. According to \cite{Gianetti17} the local variation of this ratio 
is of order \(20 \%\). Hence we adopt an overall uncertainty of $25 \%$ on the mass, not reflecting systematic uncertainties that are common to all of the clouds (which are likely much larger, but do not impact the shape of the profile).

\begin{figure}
    \centering
    \includegraphics[width = 0.45 \textwidth]{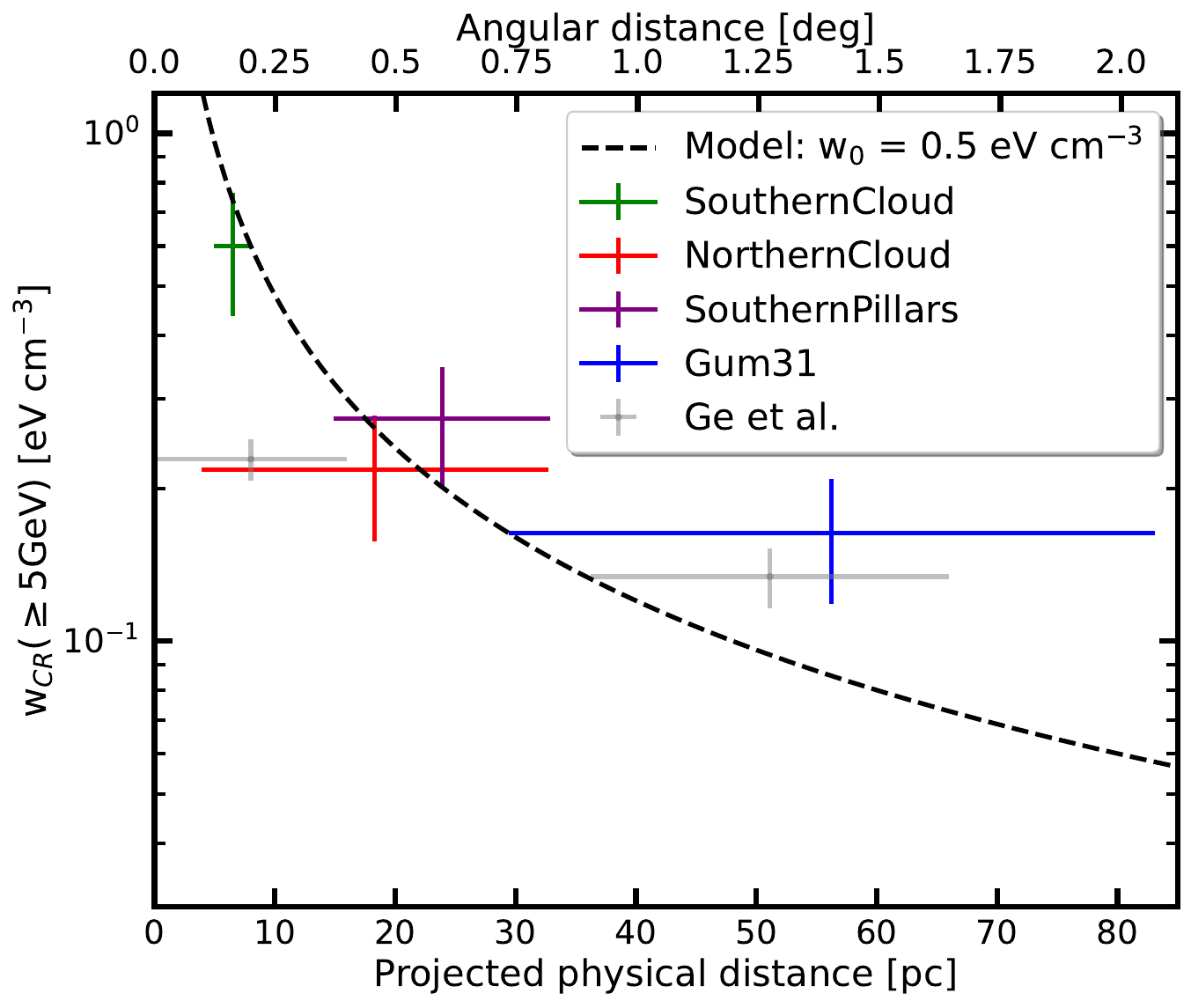}
    \caption{The CR density for each of the four clouds as a function of distance to \etacar. The angular distance has been transformed to a physical using a distance of 2.3~kpc. Distance errors depict the maximum extent of the cloud templates. A $1/r$ type profile as described in \autoref{eq:rad} has been fit to the data points and is shown by the dashed line. For comparison, the CR densities  as derived in \cite{Ge_et_al_2022_clouds_around_EtaCar} for their regions A \& B are shown in light grey.}
    \label{fig:radial_crdens}
\end{figure}

The resulting CR density profile can be seen in \autoref{fig:radial_crdens}. The physical extent is visualized by the horizontal error bars as the minimum and maximum distance from \etacar. The conversion to a physical distance scale assumes that all clouds are located in a plane at the same distance from the observer.
Compared to the result obtained by \citet{Ge_et_al_2022_clouds_around_EtaCar} the result from Gum31 is consistent with their result for region B, whereas their region A centred at \etacar\ comprises both the Southern Cloud and parts of the Northern Cloud and is hence not directly comparable.

Assuming \etacar\ as the origin of the CRs, this suggests a \(1/r\) profile, similar to that observed around some massive stellar clusters \citep{Aharonian2019},
\begin{equation}
    w_{\rm CR}(r) = w_0(r/r_o)^{-1} .
\label{eq:rad}
\end{equation}
 Normalising the profile at r$_0 = 10$~pc, a value of w$_{0} = 0.48 \pm 0.09 \UNITS{eV cm^{-3}}$ in CRs above 5~GeV is obtained from a fit to the derived CR density profile.
 Using the maximum energy flux and mass for the Homunculus, as discussed above, provides an upper limit on the CR energy density of ${\sim} 10^3 \UNITS{eV cm^{-3}}$ at a distance of $\lesssim 0.1$\,pc, which does not constrain the postulated $1/r$ behaviour. For an integration radius of 60~pc, corresponding to the outer edge of the emission seen in the CNC-Gum\,31 complex, the derived CR energy density implies a total energy of CR protons of $W_p = 5 \times 10^{48} \UNITS{erg}$. 
 
 The diffusion time can be calculated from the maximum distance, R$_{\rm max}$, that CRs have propagated for a given diffusion coefficient. Here, R$_{\rm max}$ is taken to be 60~pc and the energy-dependent diffusion is assumed to follow \autoref{eq:diff}.
Together with a total power of ${\sim}5 \times 10^{35} \UNITS{erg\,s^{-1}}$ escaping \etacar\ in CRs above an energy of 5~GeV as suggested by the model in \RW , this implies a diffusion coefficient of $D_0 = 5 \times 10^{26} \UNITS{cm^2\,s^{-1}}$. Whilst this is slower than the average Galactic diffusion coefficient ($3 \times 10^{28} \UNITS{cm^2\,s^{-1}}$), it is well above the lower limit of $5 \times 10^{25} \UNITS{cm^2\,s^{-1}}$ implied by the age of \etacar\ ($ {\sim}2 - 3 \times 10^6 $ years \citep{Mehner2010}), and so, at first sight, may appear reasonable. 

However, this diffusion coefficient implies diffusion times through the clouds that are on the same scale as the lifetime of the diffusing protons. The diffusion time through the northern cloud for our estimated diffusion coefficient and a cloud diameter of 30~pc, estimated from the size of the CO template, is ${\sim}1.4 \times 10^5 \UNITS{yr}$. The life time of the diffusing protons is approximately $t_{\rm cool} = 3 \times 10^{7} n^{-1} \UNITS{yr}$ \cite[e.g.][]{HHARAA}, corresponding to $\approx10^5 \UNITS{yr}$ for a spherical cloud of mass $10^5\, \mathrm{M}_{\odot}$ \citep{Rebolledo2015}. Consequently, the thin target approximation does not hold. Faster diffusion appears to be needed, leading to a higher total CR power which can not be produced exclusively from \etacar\ in its current state.
  
  Therefore either the CR output from \etacar\ was higher in the past by at least a factor of few, or a contribution from additional sources in the CNC is needed. \etacar\ is known to be a highly variable system on timescales of \SI{\sim 100}{\year}, and little is known about the system prior to the great eruption in the 19$^{\rm th}$ century. Equally, several candidate sources exist that could contribute to the observed emission, including massive stellar clusters (such as the nearby Trumpler~14) and massive binaries situated in the star-forming regions of the CNC \citep{Ge_et_al_2022_clouds_around_EtaCar,Aharonian2019}.

\section{Summary and conclusion}
\label{sec:discuss}
In this work, we analysed \fermi\ data from \etacar\ and four surrounding clouds and investigated CR escape, propagation and $\gamma$-ray emission from the \etacar\ binary system and its neighbourhood.
The \etacar\ system and its surroundings are extremely complex, with the wind region, the little and large Homunculus nebulae, and molecular clouds nearby. For efficient escape of particles from the system, the CRs should diffuse in the low-density region carved by the wind from \etacar-B, otherwise, they will either interact in the wind region or lose their energy via adiabatic losses. Particles diffusing into the high-density region can lead to an additional contribution to the total $\gamma$-ray spectrum. Depending on the propagation properties, the interaction of escaping particles in the Homunculus Nebula can account for a large fraction of the total $\gamma$-rays measured by \fermi.

We find that the observed emission of \etacar\ on scales not resolvable by \fermi\ can be accounted for by a variety of different models. Additionally to that put forward by \RW, this also includes models solely accelerating particles at the shock on the side \etacar-B, with the lower energy \fermi\ emission dominantly produced in the Homunculus (see \autoref{sec:prop}). Therefore, it is possible that contributions from several zones contribute to the overall emission, a fact which should be included in any future model. However, the determination of the exact amount of emission produced in each region depends on the details of the CR transport and remains a challenge.

Escaping CRs from \etacar\ may also interact in the molecular clouds of the Carina nebula. As shown in \autoref{sec:clouds}, the derived radial profile of the CR densities seems to be indicative of an origin of CRs from \etacar. However, to account for the current emission associated with molecular clouds, either \etacar\ must have been more powerful in the past, or additional CR sources must be present in the region  \cite[as proposed by][]{Ge_et_al_2022_clouds_around_EtaCar}. 

Observations of \etacar\ at several hundreds of \si{\giga\electronvolt} and \si{\tera\electronvolt} energies by Imaging Atmospheric Cherenkov Telescopes are also of great interest. This part of the spectrum is affected by absorption and potential adiabatic losses. Accurate, potentially time-dependent, observations around periastron, therefore help to investigate the propagation properties close to the stars and to constrain the emission regions. Observations from the \hess\ telescopes of the most recent periastron passage, and observations from the future CTA Observatory, may provide crucial information to help unravel the physics of this fascinating binary system \etacar.

\section{Acknowledgements}
For the numerical calculations in \autoref{sec:prop} we made use of the open-source code GAMERA \citep{gamera_2015, gamera_2022}. To estimate the shape of the escaping proton spectrum, the Naima package was used \citep{naima}.

\bibliographystyle{aa}
\bibliography{EtaCar_modelling3}

\appendix

\section{Further \fermi\ analysis details}
\label{sec:app-a}

\begin{table}[h]
    \centering
    \begin{tabular}{l|l}
    	
    	Parameter &Value \\
    	\hline
    	Data release & P8R3 \\
    	IRFs & P8R3\_SOURCE\_V3 \\
    	ROI data width & 10\grad \\
    	ROI model width & 15\grad \\
    	Bin size & 0.1\grad\ \\
    	zmax & 90\grad\ \\
    	Coordinate system & GAL \\
    	Minimum energy & 500 MeV \\
    	Maximum Energy & 500 GeV \\
    	MET start & 239557417 \\
    	MET stop & 688521600 \\
    	MET excluded (ASASSN-18fv) & 542144904 – 550885992 \\
    	evclass & 128 \\
    	evtype & 3 \\
        edisp & True \\
        edisp binning & -1 \\
        edisp disabled & Isotropic and diffuse \footnotemark \\
    	Galactic diffuse template & gll\_iem\_v07.fits \\
    	Isotropic background component & \scriptsize{iso\_P8R3\_SOURCE\_V3\_v1.txt} \\ 
    	\fermi\ catalogue & 4FGL-DR3 (\small{gll\_psc\_v29.fit}) \\
    \end{tabular}
   
    \caption{Configuration used for the \fermi\ analysis. The time period MET 542144904 – 550885992 was excluded due to the bright nova ASASSN-18fv (or V906 Carinae) \citep{Aydi20} within the ROI}.
    \label{tab:fermi_cfg}
    \end{table}

 \footnotetext{Even though disabling energy dispersion for the galactic diffuse component is not the standard procedure with a model based on the 4FGL catalog, this does not affect the result significantly due to the high energy threshold of the analysis.}
\section{$\gamma\gamma$-absorption}
\label{sec:app-b}

For the calculations of $\gamma\gamma$-absorption in the anisotropic stellar radiation fields, we used the open-source GAMERA code \citep{gamera_2015, gamera_2022}. It uses the standard cross-section for pair production which can be found, for example, in \cite{Gould_Schreder_1967}. In the form found for example in \citet[][Eq. 1]{Vernetto_Lipari_2016}, the cross-section reads:
\begin{align}
    \sigma_{\gamma\gamma} = \sigma_{\rm T} \frac{3}{16} (1- \beta^2)\left[2\beta (\beta^2 -2) + (3-\beta^4) \ln\frac{1+\beta}{1-\beta} \right],
    \label{eq:absorption}
\end{align}
where $\beta=\sqrt{1-1/x}$ with $x=s/(4 m_{\rm e}^2c^4)$ and $s=2E_{\gamma}\epsilon (1-\cos\theta)$ the square of the centre-of-mass energy. Here $\theta$ is the angle between the propagation directions of the two interacting photons, $E_{\gamma}$ the energy of the $\gamma$-ray photon, $\epsilon$ the energy of the target photon and $m_{\rm e}$ the electron rest mass.

The propagation direction of the $\gamma$-ray photon is towards the observer and the direction of the target photon depends on the position of the stars inferred from the orbital parameters. The eccentricity of the system was assumed to be \num{0.9} \citep{Damineli2008}, the semi-major axis \SI{16.64}{\astronomicalunit} \citep{Hillier2001}, the inclination of the orbit $i = \SI{135}{\degree}$ and the position angle $\phi = \SI{10}{\degree}$ \citep{Madura2012}. For each star, modelled as a black body, \autoref{eq:absorption} is integrated over the stellar surface, which is assumed to produce photons uniformly. This means that each point of the stellar surface produces the same amount of photons at the same temperature. The stellar temperatures are the same as in \SO\ and \RW, \SI{2.58e4}{\kelvin} for the primary star and \SI{3e4}{\kelvin} for the companion.

\begin{figure}[!h]
    \centering
    \includegraphics[width=0.5\textwidth]{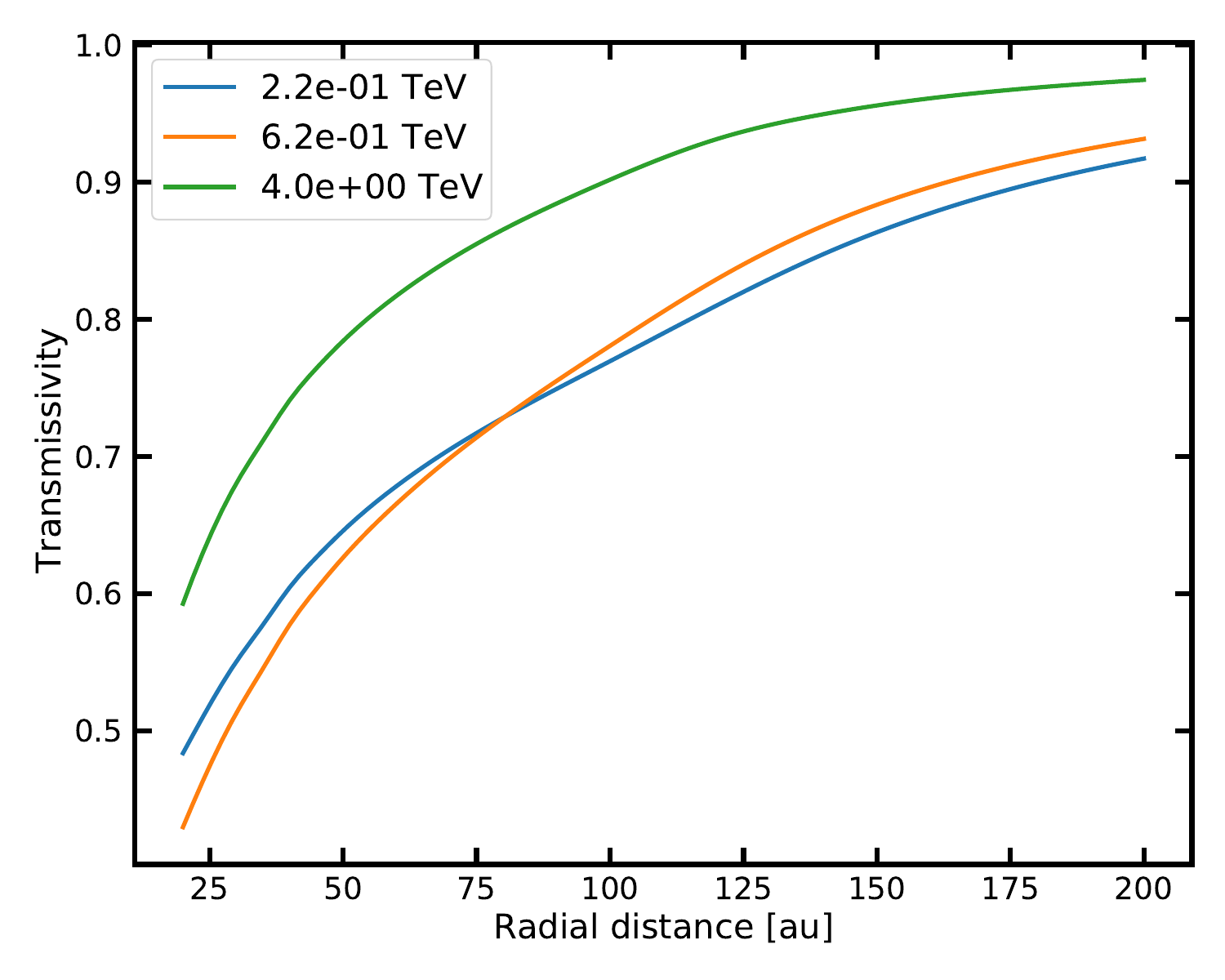}
    \caption{Transmissivity versus distance from the centre of mass of the binary system for spherical emission and different energies at an orbital phase of \num{0.5}.}
    \label{fig:absorption}
\end{figure}

\autoref{fig:absorption} shows the transmissivity at \SI{220}{\giga\electronvolt}, \SI{620}{\giga\electronvolt} and \SI{4}{\tera\electronvolt} for different distances from the centre of mass of the system assuming a spherical uniform $\gamma$-ray production. As can be seen, the transmissivity increases with increasing radius as expected due to the decreased radiation energy density at larger distances from the stars.

While at a distance of \SI{25}{\astronomicalunit}, the transmissivity at \SI{220}{\giga\electronvolt} is larger than at \SI{620}{\giga\electronvolt}, the opposite is the case for distances above \SI{\sim 75}{\astronomicalunit}. This is caused by the relative position of the two stars and their different temperatures. \etacar\ B is hotter than \etacar\ A and therefore is more important at lower energies than at higher energies. The relative absorption from each star changes with different radii because of the stellar separation, which is slightly more than \SI{30}{\astronomicalunit}, and because the centre of mass of the system is around five times more distant from \etacar\ B than from \etacar\ A. The relative absorption by target photons from \etacar\ B compared to \etacar\ A increases with distance for the apastron position creating the observed effect for the curves at \SI{220}{\giga\electronvolt} and \SI{620}{\giga\electronvolt}.

\end{document}